\documentclass{aastex631}

\usepackage{rotating}

\received{December, 2023}

\revised{March, 2024}

\accepted{May}

\submitjournal{ApJ}

\shorttitle{Rotation of Ultra-Cool Dwarfs revealed by TESS}

\shortauthors{Fontinele et al.}

\begin{document}

\title{A portrait of the rotation of Ultra-Cool Dwarfs revealed by \textit{TESS}}

\correspondingauthor{José Renan de Medeiros}
\email{renan@fisica.ufrn.br}
\author[0000-0002-3916-6441]{D. O. Fontinele}
\affiliation{Departamento de F\'isica Te\'orica e Experimental, Universidade Federal do Rio Grande do Norte, Campus Universit\'ario, Natal, RN, 59072-970, Brazil}
\author[0000-0002-7353-536X]{P. D. S. de Lima}
\affiliation{Departamento de F\'isica Te\'orica e Experimental, Universidade Federal do Rio Grande do Norte, Campus Universit\'ario, Natal, RN, 59072-970, Brazil}
\affiliation{School of Physics, Trinity College Dublin, Dublin 2, Ireland}
\author[0000-0002-2425-801X]{Y. S. Messias}
\affiliation{Departamento de F\'isica Te\'orica e Experimental, Universidade Federal do Rio Grande do Norte, Campus Universit\'ario, Natal, RN, 59072-970, Brazil}
\affiliation{Universit\'e de Montr\'eal, D\'epartement de Physique, IREX, Montr\'eal, QC H3C 3J7, Canada}
\author[0000-0002-3916-6441]{R. L. Gomes}
\affiliation{Departamento de F\'isica Te\'orica e Experimental, Universidade Federal do Rio Grande do Norte, Campus Universit\'ario, Natal, RN, 59072-970, Brazil}
\affiliation{Universit\'e de Montr\'eal, D\'epartement de Physique, IREX, Montr\'eal, QC H3C 3J7, Canada}
\author[0000-0002-3916-6441]{C. E. Ferreira Lopes}
\affiliation{Instituto de Astronom\'ia y Ciencias Planetarias, Universidad de Atacama, Copayapu 485, Copiap\'o, Chile\ }
\affiliation{Millennium Institute of Astrophysics, Nuncio Monse\~nor Sotero Sanz 100, Of. 104, 7500000 Providencia, Santiago, Chile}
\author[0000-0001-5578-7400]{B. L. Canto Martins}
\affiliation{Departamento de F\'isica Te\'orica e Experimental, Universidade Federal do Rio Grande do Norte, Campus Universit\'ario, Natal, RN, 59072-970, Brazil}
\affiliation{INAF - Osservatorio Astrofisico di Arcetri, Largo Enrico Fermi 5, 50125 Firenze, Italy}
\author[0000-0001-5845-947X]{I. C. Le\~ao}
\affiliation{Departamento de F\'isica Te\'orica e Experimental, Universidade Federal do Rio Grande do Norte, Campus Universit\'ario, Natal, RN, 59072-970, Brazil}
\author[0000-0002-3916-6441]{J. M. DE Ara\'ujo}
\affiliation{Departamento de F\'isica Te\'orica e Experimental, Universidade Federal do Rio Grande do Norte, Campus Universit\'ario, Natal, RN, 59072-970, Brazil}
\author[0000-0001-8462-4280]{E. Janot Pacheco}
\affiliation{Instituto de Astronomia, Geof\'isica e Ci\^encias Atmosf\'ericas, Universidade de S\~ao Paulo,  05509-090, São Paulo, SP, Brazil\ }
\affiliation{Observatoire de la Côte d’Azur, Laboratoire Lagrange, Bâtiment Fizeau, Parc Valrose, 06108 Nice, France}
\author[0000-0001-8218-1586]{J. R. De Medeiros}
\affiliation{Departamento de F\'isica Te\'orica e Experimental, Universidade Federal do Rio Grande do Norte, Campus Universit\'ario, Natal, RN, 59072-970, Brazil}

\begin{abstract}

This study presents the results of a search for rotation signature in 250 Gaia DR3 Ultra-Cool Dwarfs (UCDs) with TESS light curves. We identified 71 targets with unambiguous periodicities, of which 61 present rotation signatures and a single source behavior, with periods between 0.133 and 5.81 days. Five UCDs show double-dip features, namely variations with two periods, one approximately double or half the other. The remaining ten UCDs with unambiguous variability present a likely non-single behavior. We also found 20 UCDs showing complex behavior in their light curves, with noticeable fluctuations and irregular structure, with a few exhibiting apparent changes in their temporal structure. The remaining 159 targets show noisy light curves corresponding to low-amplitude signals, whose temporal variation cannot be easily identified. The distribution of the UCDs with rotation signature in the CMD diagram points to a lack of rotating objects within about $11.5<M_{G}<12.5$ and $G-G_{RP}<1.5$ separating them into two regimes, one mainly composed of less massive late-M stars with $P_{rot} \geq 1.0$ d, and another mainly composed of more massive early-M stars with $P_{rot}<1.0$,d.  It is important to emphasize that by separating stars into age intervals, one observes that UCDs with $P_{rot} \geq 1.0$\,d tend to be located in regions of younger objects, and, in contrast, those with $P_{rot}<1.0$\,d are mainly concentrated in regions of older objects. Whether  these trends of stars contrasting the sample separation is physical or produced by observational biases is a question to be verified in future studies.

\end{abstract}




\section{Introduction} \label{sec:intro}

Ultra-Cool Dwarfs (UCDs) are defined as objects with spectral types M7 or later, including L, T, and Y types, of stellar and substellar nature. These objects are characterized by intense and broad potassium lines in the optical, and water, methane, and ammonia absorption bands in the near-IR (e.g., \citealp{Kirkpatrick1999,Burgasser2002,Burningham2008}). Among their most interesting characteristics is the formation of atmospheric cloud-like structures, composed of iron and silicates for L dwarfs, chlorides and sulfides for T and early-type Y dwarfs, and water clouds in the coolest Y dwarfs (e.g., \citealp{Saumon2000,Zahnle2014,Suarez2023}). Another fundamental characteristic of the UCDs concerns their well-established variability, characterized by brightness changes caused by large-scale atmospheric structures, such as spots or longitudinal bands, which can provide fundamental information on the physical processes operating at their atmospheres (e.g., \citealp{Artigau2009,Radigan2014,Buenzli2014,Metchev2015}).

To date, the literature reports a growing list of studies on the photometric variability of UCDs, based on observations from ground-based telescopes (e.g., \citealp{TinneyTolley1999,Koen2005,Harding2013,Radigan2014,MilesPaez2017}), and from space (e.g., \citealp{Buenzli2014,Metchev2015,Cushing2016,Tannock2021,MilesPaez2023,Vos2022,Petrucci2024}). These studies revealed essential characteristics on the behavior of the periodicities, including a clustering of shortest rotation periods near 1 hr \citep{Tannock2021} and a lower envelope of rotation periodicities that runs from 2 h at spectral type M8 down to 1 h at spectral type T \citep{MilesPaez2023}. Nevertheless, the astrophysical root cause for the variability in UCDs still needs to be better understood. On the stellar side of the hydrogen-fusing limit, starspots are the usual explanation for the observed variability (e.g., \citealp{Rockenfeller2006,Irwin2011,Donati2023}). For the UCDs, the variability can have multiple origins, such as magnetic processes (e.g., \citealp{HootenHall1990}), condensated particles that form atmospheric cloud-like structures \citep{Tsuji1996}, or even brightness changes due to an atmosphere out of chemical equilibrium \citep{Tremblin2015}. \citet{Pineda2017} pointed to evidence for a complex interplay between the starspots driving the variability of UCDs and robust large-scale magnetospheric current systems.

The NASA TESS space mission \citep{Ricker2015}, launched in 2018, is producing differential photometric light curves (LCs) for hundreds of thousands of stars. Primarily dedicated to the search for terrestrial planets transiting nearby bright stars, the large number of observed targets combined with the high quality of the acquired data is opening new horizons for the study of a variety of astrophysical phenomena, including stellar rotation and activity (e.g., \citealp{Canto2020,Barraza2022,Doyle2020}). Thanks to its high-quality photometry data and continuous monitoring, TESS could also represent an optimal laboratory in the search for rotation periodicities in UCDs.

Aiming to enlarge the current sample of Ultra-Cool Dwarfs with measured rotation period, this study is dedicated to the search for rotation signatures in sources identified as Ultra-Cool Dwarfs in the Gaia DR3 \citep{Sarro2023}, observed by the TESS mission. This paper is part of a collaborative effort to determine rotation periodicities for different  families of stars located throughout the HR Diagram, with LCs collected by the referred space mission. The paper is organized as follows. Section \ref{observation} presents the stellar sample and observational data set used in this study and the procedure used to analyze the LCs. Section \ref{sec:results} provides the main results. A summary is presented in Section \ref{sec:summary}.

\section{The Ultra-Cool Dwarf Sample and Observational Data Analysis} \label{observation}

The TESS primary mission plan is to survey almost the entire sky by monitoring 26 segments (or sectors) of $90^\circ\times24^\circ$, each one with a duration of 27 days at a time, with four on-board 10.5 cm telescopes in a red (600--1000 nm) band-pass. A given sector is, therefore, revisited about every two years. However, overlaps between sectors can increase the coverage, and targets in the Continuous Viewing Zone are observed with almost unbroken coverage for a year. The mission provides photometric time series at different cadences, namely 2 and 30 minutes in Cycles 1 and 2, and 20 s, 2 minutes, and 10 minutes in Cycle 3, with a time baseline from 27 to 351 days, depending on sector overlaps. While 2-minute cadence data, also known as Target Pixel (TP) files, are available for a subset of targets, all CCDs, called full-frame images (FFIs), are read out every 30 minutes (10 minutes in the extended mission). Ten subsets of TESS targets were observed for multiple sectors, with approximately 1–2 percent of targets located in the continuous viewing zone during the primary mission \citep{Barclay2018}. Such an aspect makes these targets invaluable for extracting rotation periodicities and analyzing the persistence of stellar cycles \citep{Ferreira2015}.

For this study, searching for rotation signatures in UCDs and determining their periodicities, we have used the sample of 7630 Gaia DR3 UCDs, given by \citet{Sarro2023}, as a starting point. For the definition of this sample, those authors used the Gaia DR3 set of UCD candidates and complemented the Gaia spectrophotometry with additional photometry to characterize the candidates' global properties. This procedure includes the determination of the distances and their locus in the Gaia color-absolute magnitude diagram (CAMD). More specifically, they present the typical (median) RP spectra in spectral-type bins and compare them to UCD standards, ground-based high-resolution spectra, and CAMDs, including Gaia and external photometry. Then, that sample was compared with the TESS Input Catalogue, from which 7360 objects were identified out of the total sample of 7630. Of this latter list, we found 250 targets with TESS 2-minute cadence LCs available on October 30th, 2023. On this date, the TESS data basis also reported observations of 674 targets in a 30-min cadence, 124 in a 10-min cadence, and one target in a 20-seg cadence; all these additional data are not suitable for our purpose.

The 2-min cadence data were downloaded from the {\em FFI-TP-LC-DV Bulk Downloads Page} of the Mikulski Archive for Space Telescopes\footnote{\url{https://archive.stsci.edu/tess/bulk/downloads.html}} \citep{STScI2018} using the cURL scripts available for retrieving Pre-search Data Conditioned Simple Aperture Photometry (PDCSAP) reduced light curves. \citet{Jenkins2016} described the TESS science processing operations center (SPOC) pipeline that produces the 2-min cadence light curves. The distribution of the spectral types for the list of 250 UCDs with TESS LCs is displayed in Figure 1, which ranges from M5 to L3. The referred spectral types were computed following the calibration by \citet{2013ApJS..208....9P}. These targets are located in the following star-forming regions, open clusters, or moving groups, as listed in Table 6: Taurus (TAU), Carina-Near (CARN), Tucana-Horologium Association (THA), $\epsilon$ Chamaeleontis Association (EPSC), AB Doradus moving group (ABDMG), $\beta$ Pic Moving Group (BPMG), Upper Centaurus Lupus (UCL), Hyades (HYA), TW Hya Association (TWA), Lower Centaurus Crux (LCC), Pleiades cluster (PLE), Columba (COL), Argus (ARG), Coma Berenices(CBER), and Carina Association (Car). The determination of membership to the referred star clusters and moving groups were derived by \citet{Sarro2023}, using the BANYAN $\Sigma$ software tool \citep{gagne18}. Those authors have used the Gaia DR3 UCD candidates, namely their celestial positions and proper motions and parallaxes, as input to BANYAN $\Sigma$.

When required, additional processing on these LCs was performed manually to avoid possible distortions in the signature of periodicities. Such processing was explicitly the removal of outliers and correction of instrumental trends, following the recipe by \citet{deMedeiros2013}, \citet{PazChinchon2015}, and \citet{Canto2020}. In short, flux measurements exceeding 3$\times$ the standard deviation of the light curve compared to the average neighbor data were considered outliers and then removed from the time series. The correction of instrumental trends consisted of detrending the light curve of each sector from a polynomial fit of third order. In the analysis of the post-processed LCs for the 250 UCDs with TESS observations, we have used the same procedure applied by \citet{Canto2020}, which is based on a manifold interactive platform built in the {\em Interactive Data Language} (IDL\footnote{\url{https://www.l3harrisgeospatial.com/Software-Technology/IDL}}). Readers are referred to \citet{Canto2020} and \citet{Barraza2022} for a complete description of the referred manifold package. In short, our analysis consisted of computing Fast Fourier Transforms (FFT) (e.g., \citealp{Zhan2019}), Lomb-Scargle periodograms (e.g., \citealp{Scargle1982,Horne1986,Press1989}), as well as wavelet maps and global spectra \citep{Grossmann1984}, of each light curve. Based on the known range of the rotation periods of UCDs (e.g., \citealp{Bailer2004,Tannock2021}), we searched for the main periodicities, namely peaks, computed from those methods. The errors on the corresponding peak periods were estimated using Eq.~(2) of \citet{Lamm2004}. The Lomb-Scargle method was also used to consider the false alarm probability (FAP) of the periodicities, based on Eq. (22) of \citet{Horne1986}, given by $F = 1 - \left\{ 1 - exp[-(N_0/2)(1+xi^{-1}) - 1] \right\}^{N_i}$. The formula gives the periodogram power threshold, $F$, corresponding to a given FAP, $\xi$, where $N_0$ is the number of observations of a light curve, $\xi$ is the FAP, and $N_i$ is the number of independent frequencies, namely $N_i = -6.362 + 1.193 N_0 + 0.00098 N_0^2$, according to Eq. (13) of \citet{Horne1986}. In the present work, only peaks with FAP less than 1\%, corresponding to significance levels greater than 99\%, were taken into consideration. Additionally, wavelet maps were used to examine the persistence of the main signals within the time span of TESS observations of a given target in different sectors, and phase diagrams of the possible periodicities were computed to inspect the global morphology of each signal.

After obtaining the FFT, Lomb–Scargle, wavelet analyses, and phase diagrams  results, we visually inspected each light curve to identify the  present modulation features, adopting the procedure applied by \citet{Canto2020}.

Specifically, we separated the targets into three groups: (i) those with a typical rotational modulation in the LC, namely a signature characterized by semi-regular flux variability that used to be multi-sinusoidal; (ii) those with an ambiguous, dubious, or complex variability profile, namely those targets with LCs displaying an asymmetric shape in the flux variation, and (iii) targets with noisy LCs. From the sample of 250 UCDs with bonafide TESS LCs, listed in Table A of the Online Material, we have identified 54 UCDs presenting unambiguous variability, with variability amplitudes ranging from 0.6 to 40.9 percent, for which we have computed the associated periodicities. This list of 54 UCDs is presented in Table 1, which also contains the referred periodicities. Figure 2 displays three examples of LCs with typical rotation signatures identified in our sample of UCDs, with the corresponding phase folded LC, FFT, Lomb–Scargle periodograms, and wavelet maps. Figures following the same design as Figure 2 for all the UCDs with unambiguous variability are provided in the figure set given in the Online Material. Among the UCDs with evident variability, we identified 20 targets presenting a complex temporal structure in their LCs. The referred LCs are also provided in the figure set given in the Online Material.

\subsection{A search for periodicities in Gaia DR3 UCDs with supposedly noisy LCs}{} \label{sec:periodicities}

Our manifold analysis has initially identified 176 Gaia DR3 UCDs with LCs presenting supposedly noisy signatures, an aspect assumed particularly for LCs with a low-amplitude signal. Indeed, as underlined by different studies (e.g., \citealp{Canto2020, Gilliland2011}), a noisy behavior is a complex combination of different factors, including, for example, instrumental noise contributions and data reduction processes. Some UCDs classified as noisy LCs may have been classified this way because of data reduction issues.

Considering that different studies point to a clustering of shortest rotation periods of UCDs near 1 to 2 hr (e.g., \citealp{MilesPaez2023,Tannock2021}), we revisited the sample of 176 Gaia DR3 targets with LCs previously classified as noisy in the search for short periods, specifically for periods shorter than 1 d. First, we reanalyzed each LC using the Wavelet method to search for persistent temporal structures on time scales shorter than 1 d, all along the considered LC time span. Such a procedure revealed 17 targets with potential low amplitude variability ranging from 0.22 to 21 percent, a signature also clearly visible in the related phase folded light curve. For these targets, it was possible to compute Wavelet and LS periodicities, listed in Table 2, with the respective variability amplitude. Table 2 lists Sectors where it was possible to extract the period variability and Sectors with negligible or absent modulation. To reinforce the presence of variability with potential physical meaning, not reflecting instrumental systematics, we have analyzed the LCs of these 17 UCDs using the W Transform \citep{Wang2021,bookwang}, which computes a time-frequency as the Wavelet Transform, but with a significant difference. The W Transform uses a nonstationary window function to concentrate the spectral energy in the region of a dominant frequency and to remove the singularity problem related to the zero frequency in the time-frequency spectrum. Indeed, whereas the standard Wavelet transform shifts the energy of a signal toward a higher frequency, the W transform centralizes the referred energy around the dominant frequency. The periodicity values computed from the W Transfom are also listed in Table 2, which reinforces those obtained from Wavelet and LS methods. Figures following the same design as Figure 2 for these 17 UCDs with likely rotation signatures and low amplitude variability are provided in the figure set given in the Online Material.

 \subsection{Cleaning the sample of UCDs with modulation}{} \label{sec:source contamination}

We have checked all the stars with unambiguous periodicities to verify if the variability signals obtained from their LCs were associated with the considered UCD. Indeed, given the TESS large plate scale of 21$''$ pixel$^{-1}$, analysis of signals obtained from the LCs of this mission should be made with caution due to potential contamination by blended LCs, which can induce an interpretation of the observed variability to a wrong source \citep{Mullally2022,Higgins2023,Pedersen2023}. To assess the potential impact of contamination by nearby sources, we analyzed the crowding metric (CROWD) calculated by the SPOC pipeline \citep{Caldwell2020}, which indicates the fraction of the light in the TESS aperture that comes from the considered target given the positions and amplitudes of stars in the TESS Input Catalog. A CROWD value near 1.0 indicates a potentially isolated star, while lower values indicate a potential crowding from neighbors. This analysis found 42 UCDs with a CROWD parameter larger than 0.7 within the total sample of 71 UCDs with modulation listed in Tables 1 and 2. These tables also give the CROWD parameters, with the listed values corresponding to an average for stars observed in more than one sector. Hence, the variability amplitude of UCDs with a CROWD parameter smaller than 0.7 should be taken cautiously since these objects could be significantly affected by the flux contamination of nearby targets.

\section{The Gaia DR3 UCDs with TESS rotation periods} \label{sec:results}

Among 250 DR3  Ultra-Cool Dwarfs with TESS  bonafide available 2-min cadence LCs, we have identified 71 targets presenting unambiguous periodicities, composing two sub-samples. The first one is composed of 54 UCDs with high amplitude modulation, showing persistent temporal structures on time scales ranging from 0.133 to 5.81 days, and the second one is composed of 17 UCDs with modulation signatures despite the low amplitude of their LCs, also with persistent temporal structures on time scales shorter than one day. However, an important aspect should be underlined: a closer look into the whole of the LCs of eight of these 17 UCDs with low amplitude, namely TIC 448590406, TIC 299007548, TIC 356632694, TIC 371080717, TIC 275133493, TIC 201688405, TIC 302396416, and TIC 342967337, reveals that, in some epochs (TESS Sectors), they show a quasi-sinusoidal modulation, while in other epochs, there is a negligible or undetectable sign of variability. Tables 1 and 2  give the periodicities for the referred samples of 54 and 17 UCDs, respectively, including associated error in the period ($eP_{rot}$), effective time span ($t_{SPAN}$) of each LC (the total time span subtracted by the duration of eventual gaps), the effective number of cycles of the rotational modulation (defined as $N_{Cycle}=t_{SPAN}/P_{rot}$), and the TESS observation sectors. 

We have also identified 20 DR3 UCDs with dubious or ambiguous variability signatures in their LCs. The Dubious variability corresponds to targets showing potential rotation signature but whose period could not be disentangled among two or more possibilities (from periodogram peaks and wavelet maps) and stars with $N_{Cycle}<$ 3. Ambiguous variability corresponds to objects showing visually large-amplitude variations with a very irregular or complex structure in the LCs, noticeable fluctuations, or an insufficient time span to correctly identify the nature of their variability signature (see Section 2.5 of \citet{Canto2020}). The referred UCDs are listed in Table 2. A few of these targets, TIC 125882881, TIC 367089503, TIC 58285782, and TIC 58229208, present apparent changes in their temporal structure in the same TESS sector. 

 Tables 1 and 2 show a relatively high number of UCDs with short-period measurements, 40 percent out of the 71 objects with computed periodicities. We carefully analyzed if some of those short periods could be associated with binarity. Then, we visually inspected the short-period LCs, searching for eclipsing binary signatures. However, no sign of binarity was found in the LCs. In this sense, we have evaluated the Gaia RUWE parameter (Gaia Renormalized Unit Weight Error) for the entire list of 71 UCDs, which may be an indicator of multiplicity \citep{Lindegren2021}. A RUWE value of around 1.0 is expected for a single-star astrometric solution. In contrast, a value significantly greater than 1.0, typically greater than 1.4, could indicate a non-single source or problematic for the astrometric solution (e.g., \citealp{Kervella2022}). Ten targets with RUWE greater than 1.4 were identified, six from Table 1 (TIC 17518894, TIC 268325250, TIC 56002511, TIC 58229181, TIC 56624850, and TIC 902237947), and four from Table 2 (TIC 356632694, TIC 281668854, TIC 388682292, and TIC 448590406). The star TIC 58229181 is listed in the literature as an EB, from TESS observations \citep{Prsa2022}, whereas TIC 56624850 is classified as a circumbinary protoplanetary disk system \citep{Czekala2019}. In addition, among the UCDs with dubious or ambiguous variability signatures in their LCs, 18 present a RUWE greater than 1.4. 

 Then, for the present study, namely, the search for rotation signatures in sources identified as UCDs in the Gaia DR3, observed by the TESS mission, we consider as potential bonafide rotation signatures only those UCDs with RUWE less than 1.4. This criterion amounts to 48 and 13 objects from the lists given in Tables 1 and 2, respectively. The distribution of the computed rotation periods for the corresponding sample of 61 UCDs with RUWE less than 1.4, namely those expected to be likely single-rotating objects, is displayed in Figure 3. As it arises, the periodicities for these UCDs range from 0.133 d to 5.81 d for the sample listed in Table 1 and from 0.116 d to 0.71 d for the UCDs listed in Table 2, the latter with low LC amplitudes.

Five of the UCDs listed in Table 1 have two period values (TIC 150094175, TIC 20305594, TIC 359892714, TIC 58638214, and TIC 96680763), four of which present a second period at half the main period (TIC 20305594, TIC 359892714, TIC 58638214, and TIC96680763). These cases can be interpreted as double-dip signatures \citep{Basri2018a, Basri2018b}, typically produced by two groups of heterogeneities on the stellar surface, such as spots, clouds, or large wave bands \citep[e.g.,][]{Apai2017}, with phases shifted by about 90°. Those phase shifts may evolve along time, producing beating effects in the light curves or a variation of the peak amplitudes in the periodograms that could be caused by differential rotation \citep[e.g.,][]{Apai2017, Basri2018a, Basri2018b}. In wavelet maps, a distinctive pattern often includes two dominant features over time, where one feature's period is either double or half of the other. Frequently, the longer period corresponds to the stellar rotation period, while the shorter one results from the combination of two semi-sinusoids linked to the double-dip signature \citep[e.g.,][]{Canto2020}.

A no less relevant result concerns the high number of 159 targets showing noisy behavior in their LCs, corresponding to 64 percent of the original sample of 250 targets. Although those targets present typically low-amplitude signals whose physical periodicities cannot be easily identified, they can hide essential information. Typically, a noisy signature is a complex combination of instrumental noise contributions (related, for instance, with Poisson statistics and readout noise) plus a relevant contribution of intrinsic stellar noise, the Galactic position, light from neighboring stars, and sky background contamination (e.g., \citealp{Gilliland2011}). When the LCs considered in this work present a low-amplitude signal, we assumed them to be, in principle,  a noisy signature. The noisy behavior could reflect low activity or long periodicities for some stars, particularly for targets with short observational time spans. Table B in the Online Material gives the list of UCDs supposedly presenting noisy LCs. 

\citet{MilesPaez2023} have analyzed the TESS LCs of a sample of 23 known rapidly rotating UCDs, typically objects with a $V\sin i>30$ km/s  to explore their photometric variability. These authors used a Lomb-Scargle periodogram in their study, applying a first-order correction by removing a 24-hour median filter from the data and then a Gaussian Process regression. From such an analysis, they claimed that significant periodicities in 18 targets are compatible with rotation signature, with amplitudes ranging from 0.17 to 0.77 percent. For comparative purposes, even though only two UCDs from \citet{MilesPaez2023} are present in our Gaia DR3 UCDs whole list, we have then applied our manifold procedure to the 21 targets of their sample with 2-min cadence LCs available. The result of this analysis is presented in Table 4, from where one observes that, for most cases, our variability periods are in good agreement with those computed by  \citet{MilesPaez2023}, except for three targets (TIC 288506050, TIC 441000085, TIC 311188315), for which our analysis points instead for a noisy behavior in their LCs.  Further, it is worth underlining that among the 21 UCDs listed in Table 4, ten have a RUWE greater than 1.4.  Recently,  \citet{Petrucci2024} have claimed for rotation periodicities for a sample of 87 UCDs from TESS LCs, based on a Lomb-Scargle periodogram. Among these objects, eighty percent show $P_{rot}<1$ d. Four objects from the referred sample (TIC 298907057, TIC  902237947, TIC 201688405, and TIC 371080717) are in common with our study, showing a good agreement in the corresponding $P_{rot}$ values.

The location of the Gaia DR3 UCDs with rotation periodicities and RUWE shorter than 1.4, listed in Tables 1, 2, and 4, is shown in the color-magnitude diagram (CMD) displayed in Figure 4, from where one observes an apparent trend for a transition in the distribution of UCD rotation periods. In the left panel
UCDs with $P_{rot} \geq 1.0$ d tend to be located mainly in the regions above $11.5<M_G<12.5$, whereas those with $P_{rot}<1.0$\,d tend to populate the region below this range. In addition, the CMD diagram in the referred panel also appears to point to a dependence on the distribution of $P_{rot}$ with stellar mass, separated into two different regimes. The upper region, populated mainly by UCDs with $P_{rot} \geq 1.0$ d, is composed of lower mass M stars, whereas the lower region, mainly populated by UCDs with $P_{rot}<1.0$\,d, is composed of  larger mass M stars \citep{reyle}. Thanks to the nature of the present UCDs sample, which consists of objects located in stellar associations with measured ages (\citealp {kenton,Zuckerman06,Zuckerman11,murphy,bell,pecaut}), the right panel in Figure 4 displays the same CMD. In this panel, stars are separated into age intervals based on the values given in Tables 1 and 2, to tentatively define regions where younger and older stars are expected to dominate. The scenario emerging from the right panel in Figure 4 indicates an interesting trend: UCDs with $P_{rot} \geq 1.0$\,d tend to be located mainly in the regions of younger objects, whereas those with $P_{rot}<1.0$\,d tend to populate the regions occupied by older objects.

Nevertheless, we should be cautious with the referred trend observed in the CMD diagram, which could be affected by observational biases. The incompleteness of the analyzed UCD sample and the instrumental sensitivity in the observed wavelength band may contribute to the portrait observed in the rotation period distribution (e.g., \citealp{Canto2023,Leao2015}). For instance, the UCD sample may be incomplete to periods longer than 24 hours because TESS is mainly sensitive to L0 dwarfs and, on the other side, the lower period limits are also uncertain because most UCD campaigns have been limited to $\sim$8--24-h cadences \citep{MilesPaez2023}. Besides, the region with an apparent lack of stars with rotation period measurements could be populated by low-amplitude rotators below the TESS sensitivity. Despite these potential biases, it is worth underlining that, based on the analysis of the projected rotation velocity $V\sin i$ of UCDs, \citet{Mohanty2003} showed that this parameter tends to increase from mid-M to L types and that L UCDs are probably rapidly rotating in general, with a lack of slowly rotating in this latter spectral region. 

\section{Summary and Conclusions} \label{sec:summary}

Based on a manifold procedure, which considers the analysis of TESS LCs taking into account Lomb-Scargle periodograms, Fast Fourier Transform (FFT), and Wavelet Transform, we have analyzed a total of 250 Gaia DR3 UCDs presenting public LCs, with short-cadence TESS observations in sectors 5 to 63. We identified 48 UCDs with rotational modulation, presenting an unambiguous rotation period. Among these UCDs, five show two periods, with four presenting a second period that is approximately double or half of the rotation period. This aspect may reveal the presence of differential rotation in the referred targets. We have also identified 13 UCDs with typical modulation signatures, presenting periodicities shorter than 1 d despite a low-amplitude behavior in their LCs. Ten UCDs with unambiguous periodicities show a potential non-single behavior, pointing to a dubious origin of the referred periodicities. The present study also revealed 20 UCDs with ambiguous or dubious variability signatures in their LCs, a few of them with LCs presenting apparent changes in the temporal structure. The TESS light curves for the remaining 159 objects show a noisy pattern with very low-amplitude signals.

Considering the whole sample of UCDs with rotation periods, this study shows that 47.5 percent have a period shorter than 1 d. Another relevant result arises from the CMD diagram, which seems to show a lack of rotating UCDs within about $11.5<M_{G}<12.5$ and $G-G_{RP}<1.5$ that separates them into two different regimes. One is mainly composed of  less massive late-M stars with $P_{rot} \geq 1.0$ d, and another regime is mainly composed of  more massive early-M stars with $P_{rot}<1.0$\,d. Also, by separating the stars into age intervals, we found that UCDs with $P_{rot} \geq 1.0$\,d tend to be in regions of younger objects, while those with $P_{rot}<1.0$\,d are concentrated in regions of older objects. Whether the apparent lack of stars contrasting the sample separation is physical or produced by observational biases is a question to be verified in future studies based on a more extensive and homogeneous sample, as well as better statistics. 
If, on one side, the signatures of rotation in UCDs are associated with multiple origins, including magnetic processes, condensate particles that form atmospheric cloud-like structures, or even brightness changes due to an atmosphere out of chemical equilibrium, the objects with ambiguous variability or a noisy pattern appear to reflect also the presence of different factors, such as low activity phases or polar phenomena. In this context, the large sample of bonafide rapidly rotating UCDs revealed in the present work represents a solid laboratory to identify the cause of photometric variability in ultra UCDs through spectroscopic analysis, in particular, to identify traces that can differentiate between magnetically-induced star spots and inhomogeneous atmospheric dust clouds. A first step in this direction should be the study of their spectroscopic line profiles. If these targets are rapidly rotating, their line profiles will be significantly Doppler-broadened. The large sample of noisy UCDs also represents a relevant material for spectroscopic studies. If such behavior reflects a low activity level and a slow rotation, these objects will not show much line broadening. Identifying the root cause of photometric variability in UCDs is also mandatory to understand the structural stability of rapidly rotating Ultra Cool Dwarfs. 




\section{ACKNOWLEDGMENTS} \label{sec:ACKNOWLEDGMENTS}

Research activities of the Observational Astronomy Board at the Federal University of Rio Grande do Norte (NAOS) are supported by continuous grants from the Brazilian funding agency CNPq. This study was financed in part by the Coordena\c{c}\~ao de Aperfei\c{c}oamento de Pessoal de N\'ivel Superior $-$ Brasil (CAPES) $-$ Finance Code 001. YSM and DOF acknowledge CAPES graduate fellowships, and RLG acknowledges a CNPq PDE fellowship. PDSL and YSM acknowledge UFRN/CAPES/Print fellowship, respectively, process 88887.838338/2023-00 and 88887.717782/2022-00. JMA acknowledges the CNPq Brazilian research agency for funding (grant 311589/2021-9). BLCM, ICL, and JRM acknowledge CNPq research fellowships. This work includes data collected by the TESS mission. The NASA Explorer Program provides funding for the TESS mission. All the {\it TESS} data used in this paper can be found in MAST:\dataset[10.17909/fwdt-2x66]{http://dx.doi.org/10.17909/fwdt-2x66}. We made use of data from the European Space Agency (ESA) mission Gaia (https: //www.cosmos.esa.int/gaia), processed by the Gaia Data Processing and Analysis Consortium (DPAC, https://www.cosmos.esa.int/web/gaia/dpac/consortium). Funding for the DPAC has been provided by national institutions, particularly the institutions participating in the Gaia Multilateral Agreement. We want to thank the Referee for carefully reading the manuscript and a number of suggestions that largely improved the paper.

\begin{figure}[!ht]
	\centering
	\includegraphics[scale=.7]{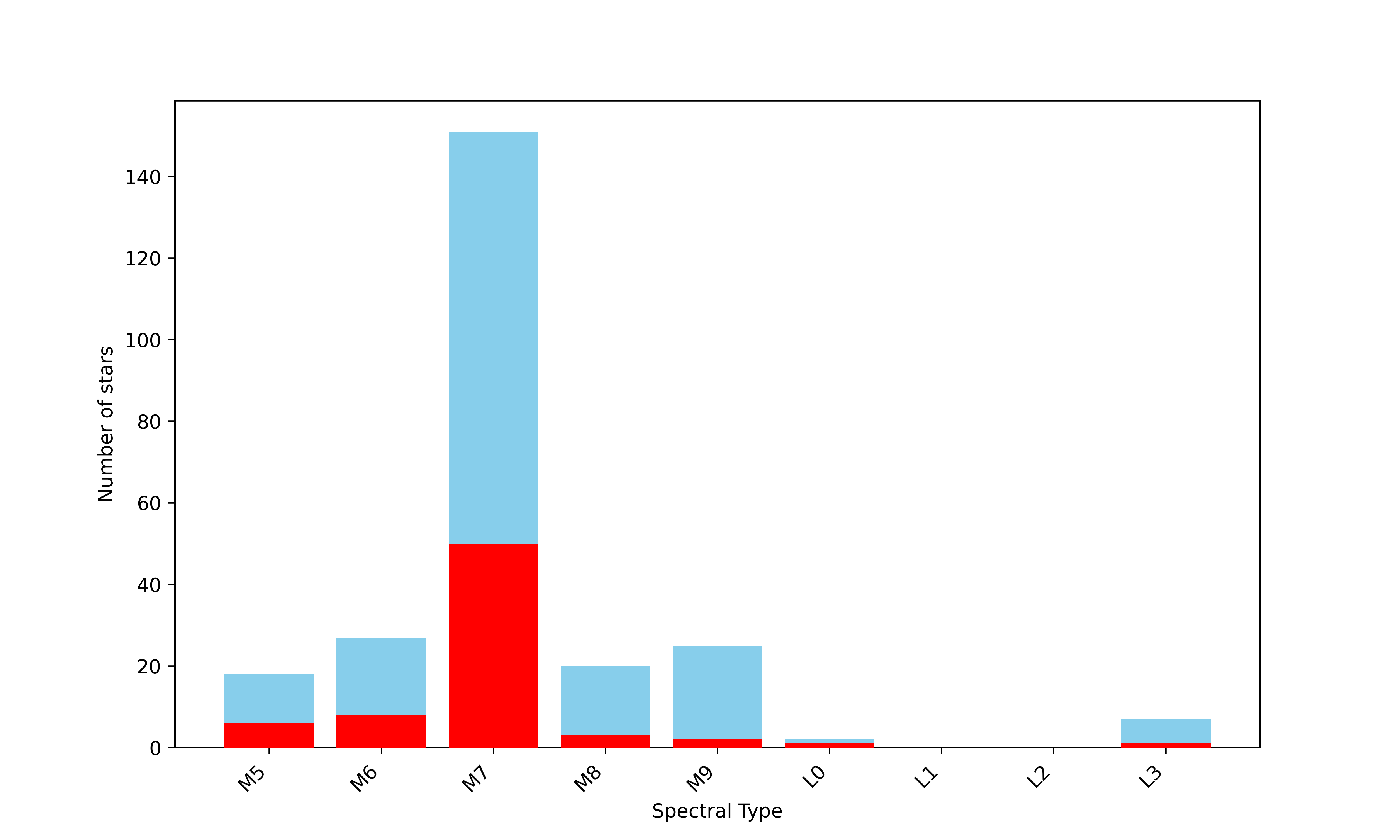}
	\caption{Distribution of spectral types for the present working sample of 250 UCs with bonafide TESS LCs. Red color stands for UCDs with unambiguous rotation and blue represents the entire UCD sample.} 
	\label{CMD}

\end{figure}

\newpage
\begin{figure*}
  \centering
  \includegraphics[width=0.60\textwidth]{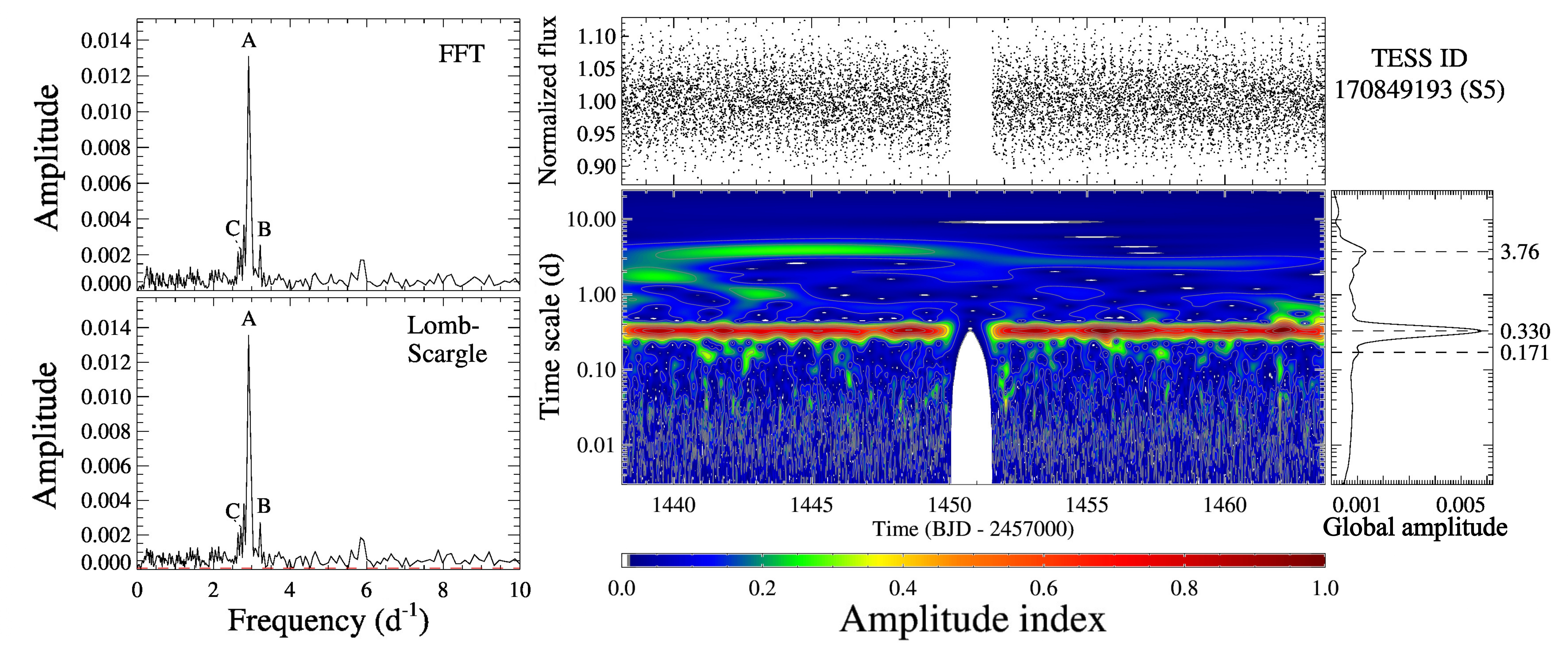}
  \includegraphics[width=0.39\textwidth]{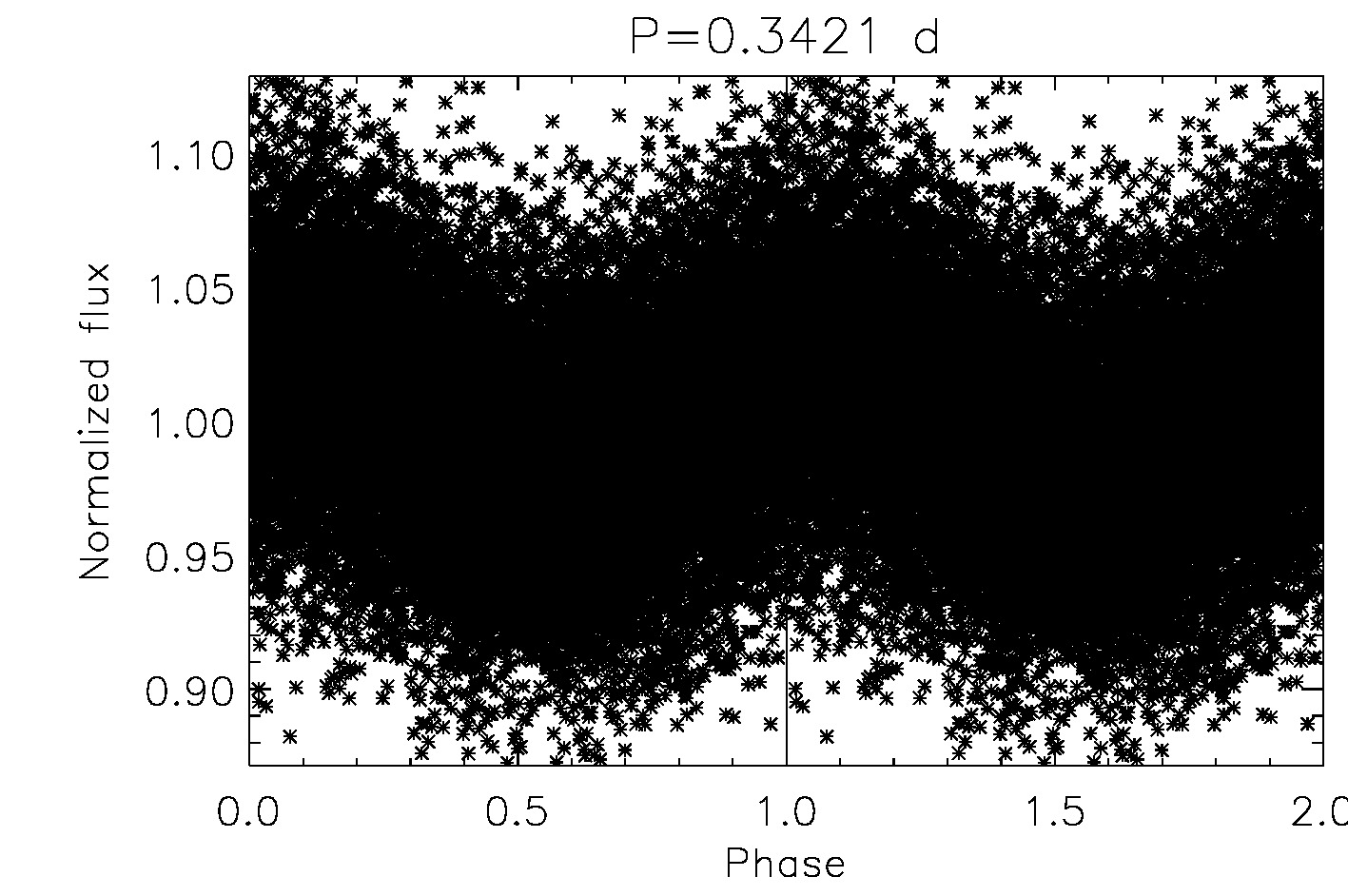} \hspace{1cm}
  \includegraphics[width=0.60\textwidth]{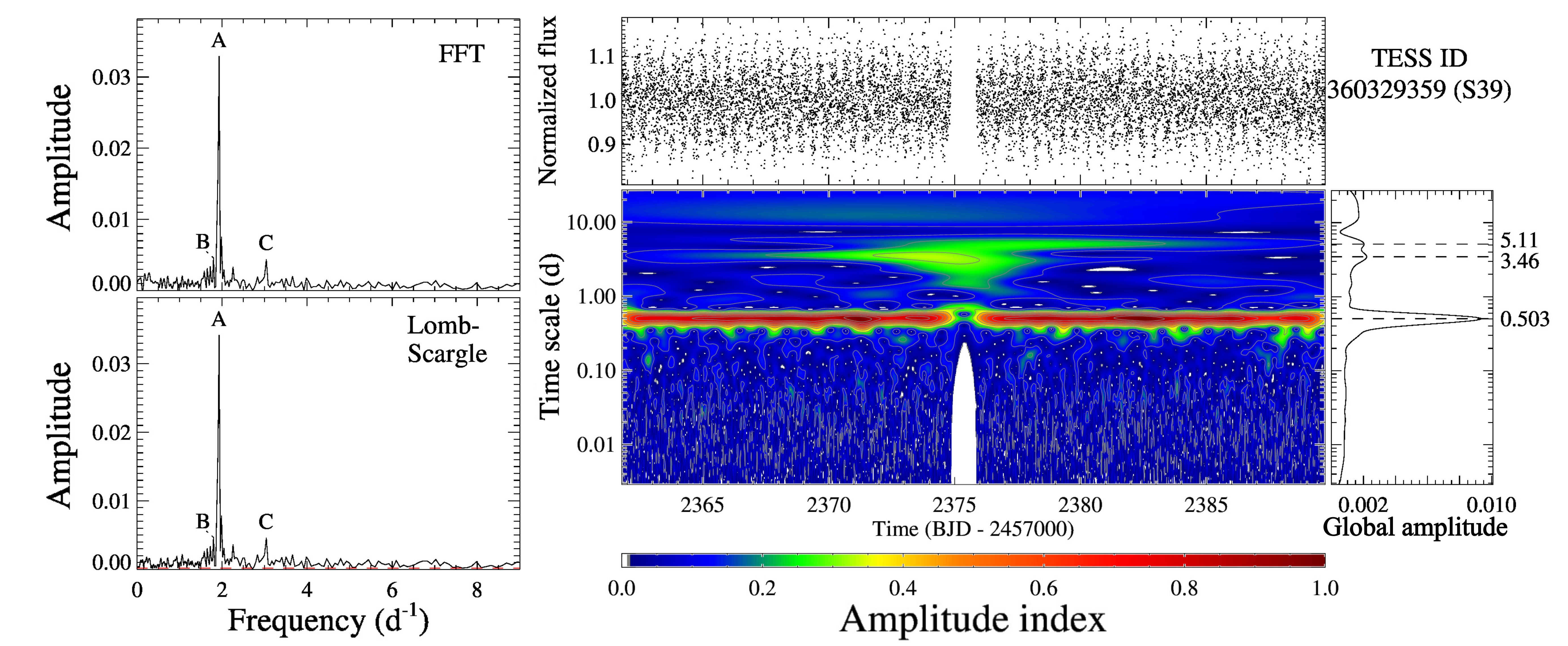}
  \includegraphics[width=0.39\textwidth]{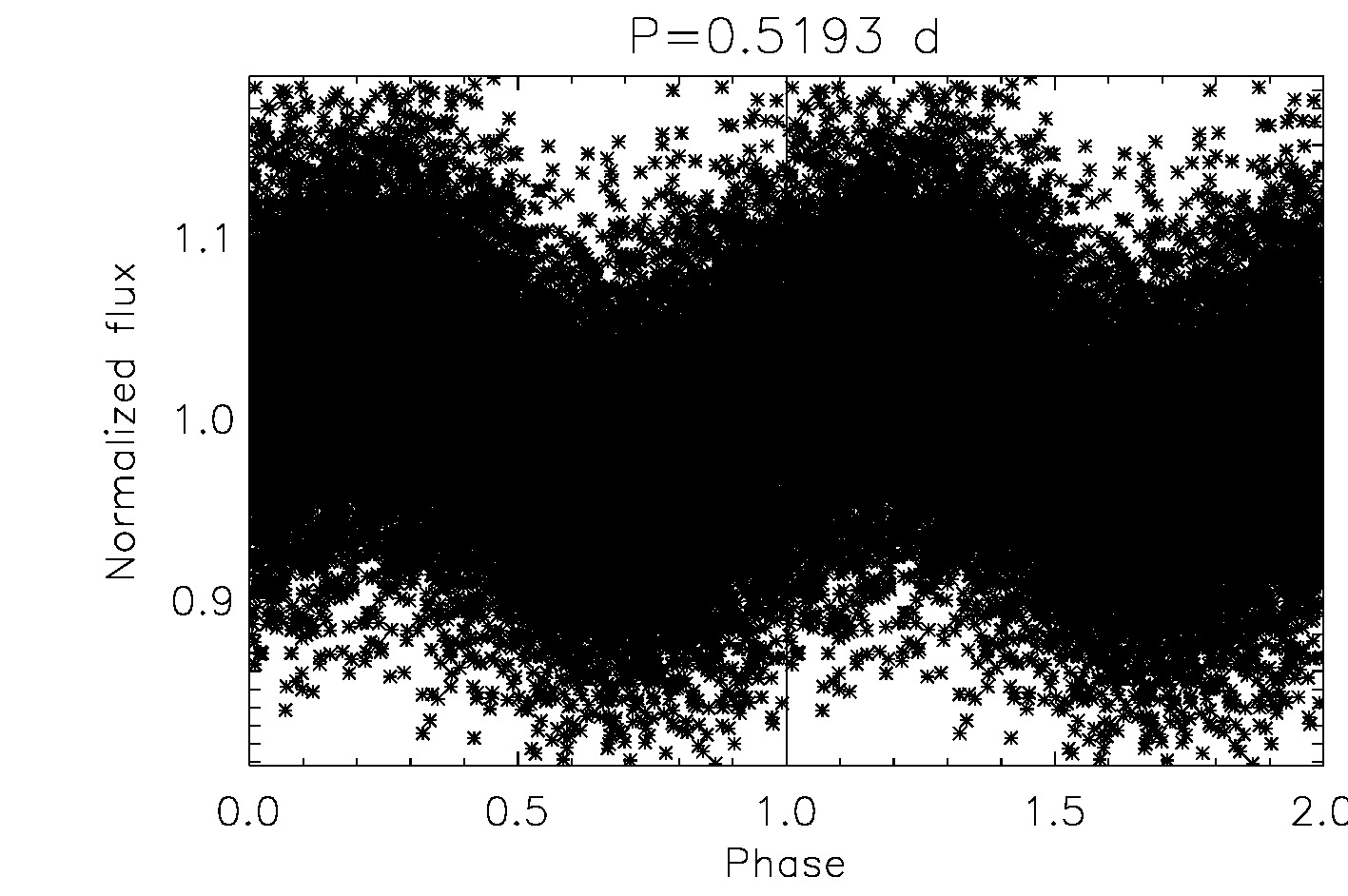}\hspace{1cm}
  \includegraphics[width=0.60\textwidth]{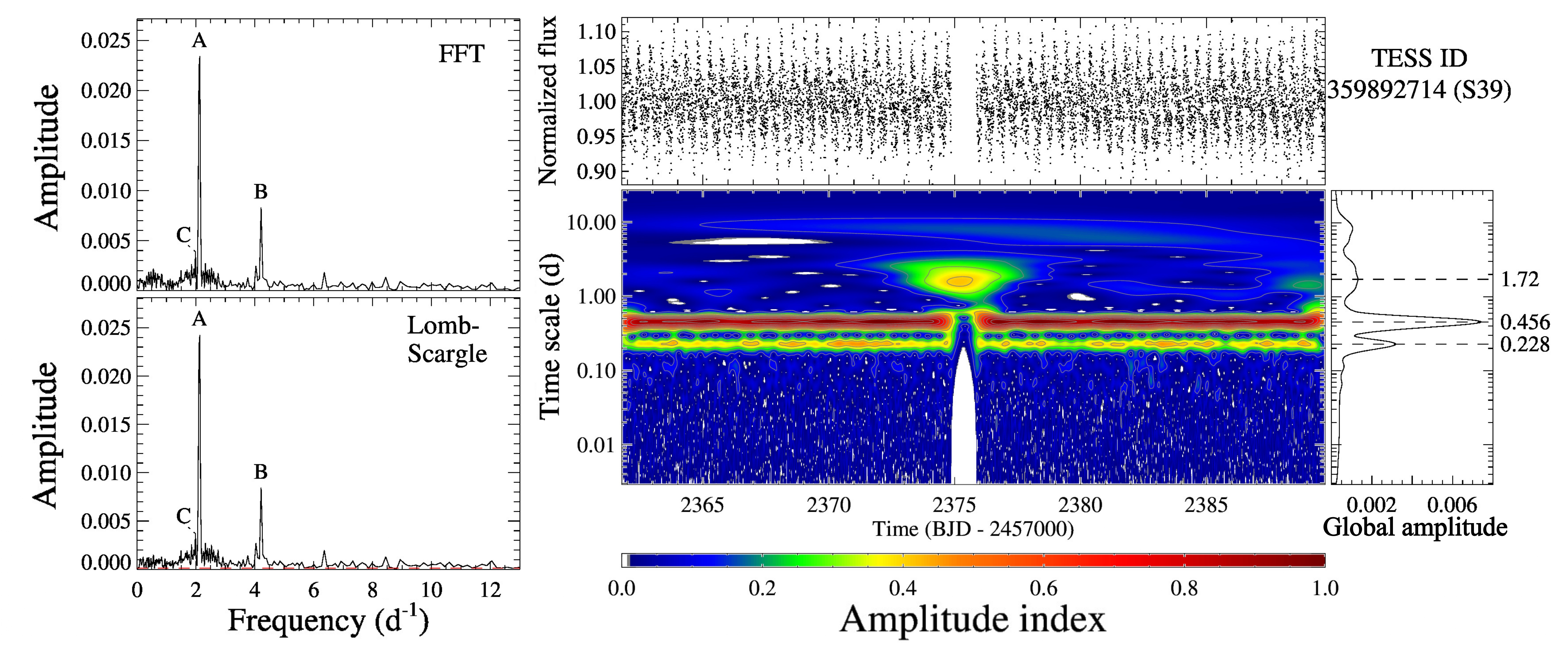}
  \includegraphics[width=0.39\textwidth]{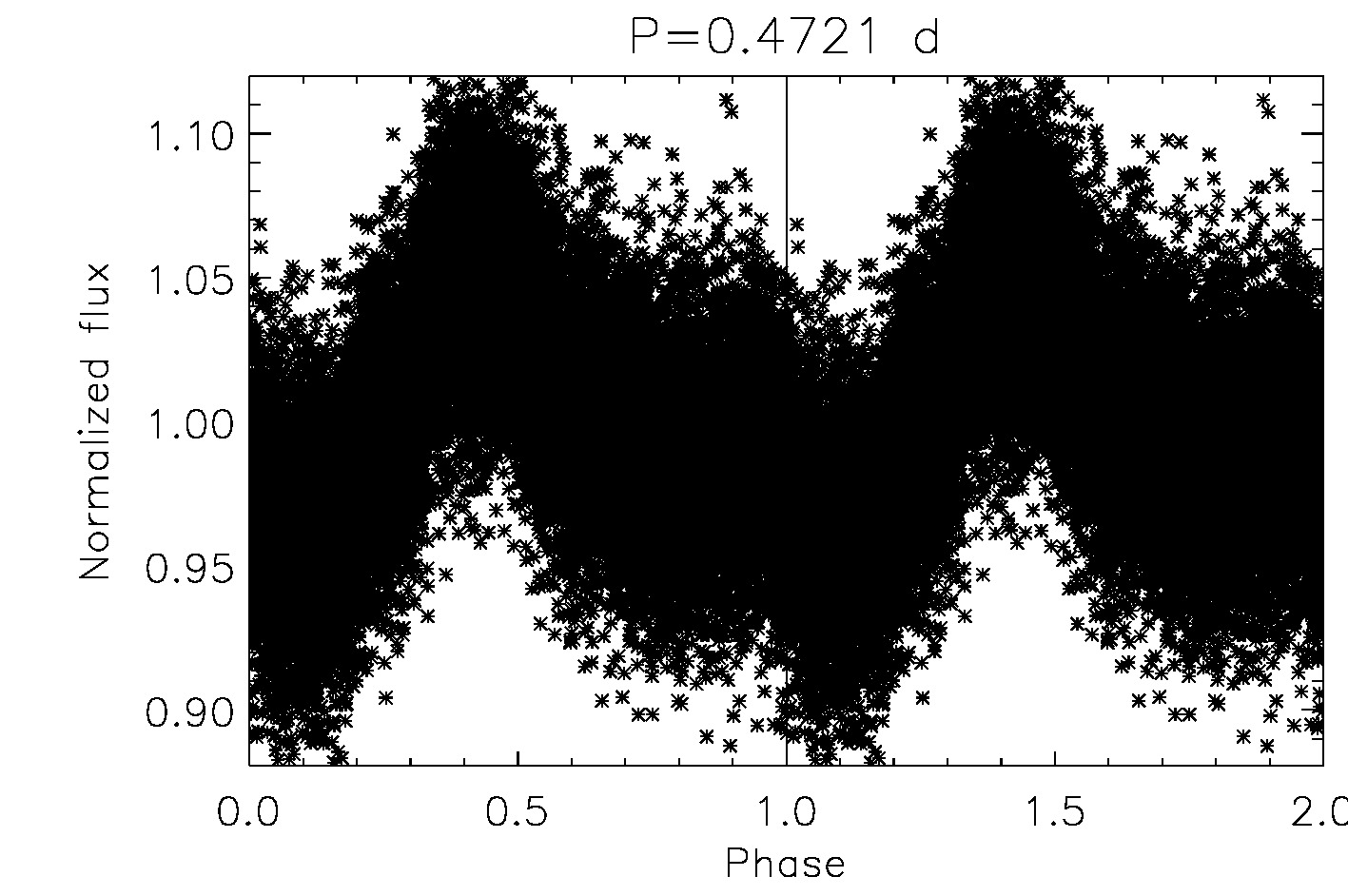}
  
   \caption{Examples of diagnostic plots displaying FFT and Lomb–Scargle periodograms, LCs, LCs phase folded, and wavelet maps for three Gaia DR3  Ultra-Cool Dwarfs with typical TESS rotation signatures.
   Persistent variability of rotation period 0.329,  0.519 and 0.472 days, respectively, for TIC 170849193 (top panels), TIC 360329359 (middle panels), and TIC 359892714 (bottom panels) are observed in their wavelet maps and confirmed by FFT and Lomb–Scargle peaks labeled A. Figures for the entire sample, following the same design, are provided in the figure set given in the Online Material}
  \label{fig_CorotLCs}
\end{figure*}
{\tabletypesize{\scriptsize}
\startlongtable
\begin{deluxetable*}{c c c c l c c c c c }
\tablenum{1}
\tablecaption{Gaia DR3  Ultra-Cool Dwarfs with TESS unambiguous periodicities\label{tab:messier}}
\tablewidth{0pt}
\tablehead{
\colhead{TIC ID} & \colhead{GAIA ID} &\colhead{P$_{rot}$} & \colhead{eP$_{rot}$} &
\colhead{t$_{SPAN}$} & \colhead{N$_{Cycle}$} & \colhead{Sectors} &  \colhead{CROWD} & \colhead{RUWE}& \colhead{Age}\\
\colhead{} & \colhead{} & \colhead{(days)} &\colhead{(days)} &\colhead{(days)} & \colhead{} & \colhead{} &  \colhead{}  & \colhead{} & \colhead{(Myr)}} 

\startdata
5630425		&	3493736924979792768	&	1.499	&	0.018	&	21.000	&	14.000	&	10	&	0.494	&	1.104	&	24\tablenotemark{\scriptsize{a}}	\\
17308621		&	48193316928788352	&	1.332	&	0.020	&	43.981	&	33.019	&	43,44	&	0.589	&	1.357	&	2\tablenotemark{\scriptsize{b}}	\\
17518894		&	3314586142482134144	&	2.065	&	0.049	&	43.916	&	21.267	&	43,44	&	0.988	&	1.491	&	2\tablenotemark{\scriptsize{b}}	\\
20305594		&	1388684770725399296	&	0.281	&	0.001	&	73.210	&	260.534	&	23,24,50,51	&	0.994	&	1.183	&	149\tablenotemark{\scriptsize{a}}	\\
45193132		&	5314960632252986880	&	2.486	&	0.025	&	123.355	&	49.620	&	8-10,35-37	&	0.093	&	1.01	&	149\tablenotemark{\scriptsize{a}}	\\
49178529		&	5401822669314874240	&	0.967	&	0.022	&	21.503	&	22.237	&	36	&	0.436	&	1.085	&	10\tablenotemark{\scriptsize{a}}	\\
56002511		&	3185321611286286976	&	0.377	&	0.003	&	23.785	&	63.089	&	5	&	0.956	&	2.317	&	149\tablenotemark{\scriptsize{a}}	\\
56624850		&	163178353176600448	&	3.155	&	0.113	&	44.177	&	14.002	&	43,44	&	0.782	&	1.551	&	149\tablenotemark{\scriptsize{a}}	\\
56658216		&	164802984685384320	&	2.606	&	0.077	&	44.140	&	16.938	&	43,44	&	0.823	&	1.013	&	2\tablenotemark{\scriptsize{b}}	\\
56658273		&	164800235906367232	&	5.81	&	0.382	&	44.160	&	7.601	&	43,44	&	0.894	&	0.888	&	2\tablenotemark{\scriptsize{b}}	\\
58229181		&	164513022853468160	&	2.11	&	0.050	&	44.140	&	20.919	&	43,44	&	0.978	&	1.54	&	2\tablenotemark{\scriptsize{b}}	\\
58285779		&	164513400810646912	&	1.669	&	0.063	&	22.000	&	13.182	&	44	&	0.183	&	1.205	&	2\tablenotemark{\scriptsize{b}}	\\
58436495		&	150501362066641664	&	0.58	&	0.004	&	43.885	&	75.664	&	43,44	&	0.270	&	0.979	&	2\tablenotemark{\scriptsize{b}}	\\
58538719		&	152516079683687680	&	2.447	&	0.068	&	44.046	&	18.000	&	43,44	&	0.872	&	1.036	&	2\tablenotemark{\scriptsize{b}}	\\
58601455		&	152108882425024128	&	4.66	&	0.251	&	43.211	&	9.273	&	44	&	0.568	&	0.976	&	2\tablenotemark{\scriptsize{b}}	\\
58638214		&	152917298349085824	&	0.752	&	0.004	&	67.494	&	89.753	&	19,43,44	&	0.726	&	1.372	&	2\tablenotemark{\scriptsize{b}}	\\
78772203		&	5424690587034982144	&	0.133	&	0.000	&	88.644	&	666.499	&	35,36,62,63	&	0.173	&	1.08	&	40\tablenotemark{\scriptsize{c}}	\\
96680763		&	156915878541979008	&	0.302	&	0.001	&	44.205	&	146.423	&	43,44	&	0.107	&	0.928	&	2\tablenotemark{\scriptsize{b}}	\\
101256057		&	216677698470254976	&	1.691	&	0.023	&	61.735	&	36.493	&	42,43,44	&	0.017	&	0.87	&	$-$		\\
102076870		&	3459372646830687104	&	0.982	&	0.011	&	41.944	&	42.713	&	10,37	&	0.412	&	1.068	&	10\tablenotemark{\scriptsize{a}}	\\
118520413		&	145238687096970496	&	1.498	&	0.026	&	43.212	&	28.846	&	43,44	&	0.894	&	0.981	&	2\tablenotemark{\scriptsize{b}}	\\
125882806		&	148420639387738112	&	1.027	&	0.012	&	43.514	&	42.342	&	43,44	&	0.980	&	0.944	&	2\tablenotemark{\scriptsize{b}}	\\
125977598		&	147441558642852736	&	4.418	&	0.223	&	43.794	&	9.913	&	43,44	&	0.970	&	1.001	&	2\tablenotemark{\scriptsize{b}}	\\
150005829		&	148354733113981696 	&	3.138	&	0.113	&	43.534	&	13.873	&	43,44	&	0.895	&	1.285	&	2\tablenotemark{\scriptsize{b}}	\\
150058659		&	148400229703257856	&	3.438	&	0.136	&	43.566	&	12.672	&	43,44	&	0.980	&	1.021	&	2\tablenotemark{\scriptsize{b}}	\\
150094175		&	157644373715415424	&	0.726	&	0.006	&	43.727	&	60.230	&	43,44	&	0.674	&	1.109	&	2\tablenotemark{\scriptsize{b}}	\\
150122702		&	148196510814073728	&	3.782	&	0.164	&	43.681	&	11.550	&	43,44	&	0.894	&	0.969	&	2\tablenotemark{\scriptsize{b}}	\\
167808464		&	5286310069347838080 	&	0.279	&	0.000	&	145.412	&	583.983	&	7-13	&	0.223	&	0.99	&	45\tablenotemark{\scriptsize{a}}	\\
170849193		&	4877824564574114304	&	0.329	&	0.002	&	23.768	&	72.243	&	5	&	0.845	&	1.079	&	45\tablenotemark{\scriptsize{a}}	\\
232064183		&	6406967509044866560	&	0.718	&	0.006	&	41.292	&	57.510	&	27,28	&	0.064	&	1.062	&	24\tablenotemark{\scriptsize{a}}	\\
268017136		&	151262876946558976	&	2.78	&	0.088	&	43.860	&	15.777	&	43,44	&	0.767	&	0.995	&	2\tablenotemark{\scriptsize{b}}	\\
268018471		&	152466120624336896	&	2.037	&	0.047	&	43.914	&	21.558	&	43,44	&	0.879	&	1.08	&	2\tablenotemark{\scriptsize{b}}	\\
268144659		&	152643240779301632	&	1.926	&	0.042	&	43.84	&	22.762	&	43,44	&	0.287	&	1.033	&	2\tablenotemark{\scriptsize{b}}	\\
268148996		&	 151327159721125888 	&	1.389	&	0.022	&	43.38	&	31.231	&	 43,44	&	0.885	&	1.133	&	2\tablenotemark{\scriptsize{b}}	\\
268217504		&	149369139966814976	&	3.918	&	0.176	&	43.553	&	11.116	&	43,44	&	0.966	&	1.095	&	2\tablenotemark{\scriptsize{b}}	\\
268218180		&	146366442430208640	&	2.382	&	0.065	&	43.528	&	18.274	&	43,44	&	0.976	&	1.128	&	2\tablenotemark{\scriptsize{b}}	\\
268324394		&	146277553787186048	&	3.379	&	0.132	&	43.310	&	12.817	&	43,44	&	0.555	&	1.049	&	2\tablenotemark{\scriptsize{b}}	\\
268324578		&	147799209159857280 	&	2.248	&	0.058	&	43.346	&	19.282	&	43,44	&	0.142	&	1.213	&	2\tablenotemark{\scriptsize{b}}	\\
268325250		&	151028990206478080	&	2.386	&	0.066	&	43.273	&	18.136	&	43,44	&	0.723	&	1.5	&	2\tablenotemark{\scriptsize{b}}	\\
268397898		&	147801339463632000	&	1.004	&	0.012	&	43.282	&	43.110	&	43,44	&	0.987	&	1.353	&	2\tablenotemark{\scriptsize{b}}	\\
268399069		&	151130591952773632	&	1.866	&	0.040	&	43.456	&	23.288	&	43,44	&	0.952	&	1.249	&	2\tablenotemark{\scriptsize{b}}	\\
298907057		&	3200303384927512960	&	0.484	&	0.005	&	23.879	&	49.337	&	5,32	&	0.825	&	1.099	&	200\tablenotemark{\scriptsize{d}}	\\
359892714		&	5841324542416019712	&	0.228	&	0.000	&	97.523	&	427.733	&	11,12,38,39	&	0.669	&	1.069	&	3.7	\tablenotemark{\scriptsize{e}}	\\
360329359		&	5919792529769594240	&	0.502	&	0.005	&	26.857	&	53.500	&	39	&	0.142	&	0.974	&	149\tablenotemark{\scriptsize{a}}	\\
367079252		&	2278720295036404736	&	2.769	&	0.029	&	132.993	&	48.029	&	17-19,24-26	&	0.906	&	1.026	&	$-$		\\
379097589		&	3724596914697213056	&	0.237	&	0.001	&	33.159	&	139.910	&	23,50	&	0.606	&	1.146	&	200\tablenotemark{\scriptsize{d}}	\\
397287296		&	3314309890186259712	&	2.655	&	0.080	&	43.785	&	16.492	&	43,44	&	0.826	&	1.125	&	2\tablenotemark{\scriptsize{b}}	\\
401838575		&	4954453580066220800	&	0.315	&	0.001	&	43.106	&	136.844	&	2,3	&	0.978	&	1.178	&	45\tablenotemark{\scriptsize{a}}	\\
427770178		&	147614422487144960	&	1.479	&	0.025	&	43.355	&	29.314	&	43,44	&	0.726	&	1.198	&	2\tablenotemark{\scriptsize{b}}	\\
440859813		&	6224387727748521344	&	2.04	&	0.047	&	44.361	&	21.746	&	11,38	&	0.480	&	1.256	&	149\tablenotemark{\scriptsize{a}}	\\
454227159		&	5201185574182015744	&	2.042	&	0.020	&	51.000	&	25.07	&	38,39	&	0.395	&	1.145	&	2\tablenotemark{\scriptsize{b}}	\\
454291779		&	5201129563515179520	&	3.23	&	0.075	&	70.000	&	21.672	&	11,38,39	&	0.848	&	1.024	&	$-$		\\
456944264		&	3314299238667410176	&	2.22	&	0.056	&	43.834	&	19.745	&	43,44	&	0.898	&	1.048	&	2\tablenotemark{\scriptsize{b}}	\\
902237947		&	3562157781229213312	&	0.241	&	0.001	&	21.439	&	88.884	&	36	&	0.210	&	3.117	&	200\tablenotemark{\scriptsize{d}}	\\
        &                     &       &       &        &         &                \\
\hline
\multicolumn{8}{c}{Targets with 2nd period}                                                          \\
\hline
          &                     &       &       &        &         &                \\
20305594	&	1388684770725399296	&	0.563	&	0.002	&	73.210	&	130.035	&	23,24,50,51	&	0.994	&	1.183 &	149\tablenotemark{\scriptsize{a}}	\\
58638214	&	152917298349085824	&	0.367	&	0.001	&	67.494	&	183.758	&	19,43,44	&	0.726	&	1.372 &	2\tablenotemark{\scriptsize{b}} \\
96680763	&	156915878541979008	&	0.117	&	0.000	&	44.205	&	377.821	&	43,44	&	0.107	&	0.928&	2\tablenotemark{\scriptsize{b}} \\
150094175	&	157644373715415424	&	0.966	&	0.011	&	43.727	&	45.266	&	43,44	&	0.674	&	1.109	&	2\tablenotemark{\scriptsize{b}} \\
359892714	&	5841324542416019712	&	0.458	&	0.001	&	97.523	&	212.932	&	11,12,38,39	&	0.669	&	1.069&	3.7\tablenotemark{\scriptsize{e}}\\
\enddata
\tablecomments{ Ages are from  (\textit{a})\citep{bell}, (\textit{b})\citep{kenton}, (\textit{c})\citep{Zuckerman11}, (\textit{d})\citep{Zuckerman06}, (\textit{e})\citep{murphy}}
\end{deluxetable*}}

{\tabletypesize{\scriptsize}
\startlongtable
\begin{deluxetable*}{cccccccccc}
\tablenum{2}
\tablecaption{Gaia DR3  Ultra-Cool Dwarfs with low-amplitude and TESS unambiguous periodicities.}
\tablewidth{0pt}
\tablehead{
\colhead{TIC ID} & \colhead{GAIA ID}  &   \colhead{P$_{rot}$ (LS)} & 
\colhead{Amplitude} & \colhead{P$_{rot}$ (WT)}& \colhead{TESS sectors } & \colhead{Noisy}& \colhead{CROWD} & \colhead{RUWE} & \colhead{Age}\\
\colhead{} & \colhead{}     & \colhead{(days)} &\colhead{(\%)} & \colhead{(days)} &\colhead{with modulation} & \colhead{TESS sectors} & \colhead{}& \colhead{} & \colhead{(Myr)}}
\startdata
26126812	&	 68012529415816832 	&	 0.229 	&	 0.50	&	0.260	&	 42,43,44 	&		&	0.974	&	 1.217	&	24\tablenotemark{\scriptsize{a}}	\\
50085560	&	 2613754712222934656 	&	 0.427 	&	 0.80	&	0.444	&	 42 	&		&	0.978	&	 0.958	&	149\tablenotemark{\scriptsize{a}}	\\
52256020	&	4691426694779173376	&	0.265	&	0.50	&	0.254	&	1,2,27-29	&		&	0.971	&	 1.292	&	45\tablenotemark{\scriptsize{a}}	\\
58432630	&	152284735566828032	&	0.642	&	0.85	&	0.690	&	43,44	&		&	0.511	&	 0.945	&	2\tablenotemark{\scriptsize{b}}	\\
201688405	&	 6473160651658069120	&	0.692	&	0.39	&	0.731	&	27	&	13	&	0.850	&	 0.983	&	149\tablenotemark{\scriptsize{a}}	\\
266012525	&	 2556491429388743296 	&	 0.313 	&	 1.76 	&	0.313	&	 43 	&		&	0.971	&	 1.079	&	45\tablenotemark{\scriptsize{a}}	\\
275133493	&	 3852362605385919616 	&	 0.438 	&	 0.73	&	0.446	&	8,35	&	45	&	0.808	&	 1.042	&	149\tablenotemark{\scriptsize{a}}	\\
281668854	&	 4920330653311504256 	&	0.116	&	 0.56	&	0.118	&	 2,28,29 	&		&	0.877	&	 9.984	&	45\tablenotemark{\scriptsize{a}}	\\
299007548	&	 3230008650057256960 	&	 0.529 	&	 0.40	&	0.568	&	32	&	5	&	0.809	&	 1.140	&	24\tablenotemark{\scriptsize{a}}	\\
302396416	&	 1113083282052337792	&	0.458	&	0.42	&	0.464	&	19,20,26,40,47	&	60,53	&	0.832	&	 1.020	&	149\tablenotemark{\scriptsize{a}}	\\
342967337	&	 5316349173696052736	&	0.705	&	2.62	&	0.697	&	8	&	9,10	&	0.126	&	 1.056	&	45\tablenotemark{\scriptsize{a}}	\\
356632694	&	 1640572714166158976 	&	 0.240	&	 0.27 	&	0.260	&	 14,16,21,49 	&	15,22-24,41,48,50	&	0.705	&	 1.426	&	149	\tablenotemark{\scriptsize{a}}	\\
371080717	&	 898275195732381440 	&	 0.714 	&	 0.35	&	0.714	&	20	&	60	&	0.311	&	 1.057	&	40\tablenotemark{\scriptsize{c}}	\\
388682292	&	 5514929155583865216 	&	 0.533 	&	 9.80	&	0.543	&	 8 	&		&	0.099	&	 2.109	&	149\tablenotemark{\scriptsize{a}}	\\
405514195	&	 6059992460039778560 	&	 0.237 	&	 2.90	&	0.240	&	 11 	&		&	0.160	&	 1.047	&	15\tablenotemark{\scriptsize{d}}	\\
448590406	&	 3478519134297202560 	&	 0.404 	&	 0.16 	&	0.431	&	 36 	&	10	&	0.769	&	 2.201	&	10\tablenotemark{\scriptsize{a}}	\\
959690391	&	 5856405272135505024 	&	 0.221 	&	 21.69 	&	0.223	&	 38 	&		&	0.005	&	 1.144	&	149\tablenotemark{\scriptsize{a}}	\\
\enddata
\tablecomments{ Ages are from  (\textit{a})\citep{bell}, (\textit{b})\citep{kenton}, (\textit{c})\citep{Zuckerman11}, (\textit{d})\citep{pecaut}}
\end{deluxetable*}
}

\startlongtable
\begin{deluxetable*}{ccccc}
\tablenum{3}
\tablecaption{Gaia DR3  Ultra-Cool Dwarfs with TESS ambiguous or dubious variability.\label{Ambiguous or Dubious}}
\tablewidth{0pt}
\tablehead{
\colhead{TIC ID} & \colhead{GAIA ID} & \colhead{Ambiguous or Dubious } 
 & \colhead{Noisy} & \colhead{RUWE}\\
\colhead{} & \colhead{} & \colhead{TESS sectors} &\colhead{TESS sectors} & \colhead{}
}
\startdata
26390890	&	216704503361774080	&	42,43,44	&		&	1.057	\\
58108762	&	164474986623118592	&	43,44	&		&	0.880	\\
58175278	&	164783811951433856	&	43,44	&		&	3.177	\\
58229208	&	164702070133970944	&	19,43,44	&		&	1.216	\\
58285782	&	164502062096975744	&	19,43,44	&		&	1.408	\\
58544613	&	149629483705467008	&	43,44	&		&	1.152	\\
58601266	&	152029992465874560	&	43,44	&		&	1.024	\\
118710667	&	146874275068113664	&	43,44	&		&	0.958	\\
125882881	&	148449845165337600	&	43,44	&		&	1.080	\\
167670065	&	6118581861234228352	&	11	&	38	&	0.974	\\
260304085	&	 5496160698257650048	&	4	&	5,7,8,10-13	&	0.991	\\
346721695	&	 216575791780899200	&	42,44	&	43	&	1.004	\\
348844681	&	 5479241065435684352	&	13	&	5,7,8,12	&	0.987	\\
367089503	&	2279554205887675392	&	17,19,24,25	&		&	1.149	\\
385557215	&	71504509626338048	&	42	&	43,44	&	0.909	\\
387813466	&	2270043769607851264	&	16, 17,18,24, 25	&		&	1.135	\\
389053345	&	3311785239689504512	&	44	&	43	&	1.020	\\
397367343	&	3480771277705948928	&	10	&		&	1.041	\\
436574130	&	3406128761895775872	&	43,44	&		&	0.964	\\
467753980	&	 169163647804297216	&	44	&	43	&	1.010	\\
\enddata

\end{deluxetable*}

\newpage
{\tabletypesize{\tiny}
\begin{deluxetable*}{c|cccc|ccc|cc}
\tablenum{4}
\tablecaption{Gaia DR3  Ultra-Cool Dwarfs with previous TESS periodicities in the literature.}
\tablehead{
\colhead{TIC ID}  &   \colhead{P$_{rot}$(h)} & \colhead{Amplitude(\%)} & \colhead{TESS sectors} & \colhead{TESS sectors}& \colhead{P$_{rot}$(h)}&\colhead{Amplitude(\%)} &\colhead{TESS sectors}  &\colhead{CROWD} &\colhead{RUWE}\\
 \colhead{} &\multicolumn{2}{c}{this work}  &\colhead{ with modulation-this work} & \colhead{Noisy-this work}  &\multicolumn{3}{c}{Miles-Páez et al. (2022)}&\colhead{} &\colhead{} }
\startdata
8259901	&		3.072 $\pm$ 0.007				&		0.93		&		21		&		—		&		2.5267 $\pm$ 0.0002		&		0.44$\pm$0.01		&		21		&		0.240	&		2.785		\\
8259902	&		3.072 $\pm$ 0.005				&		0.26		&		21,48		&		—		&		3.0889 $\pm$ 0.0002		&		0.44$\pm$0.01		&		21		&		0.829	&		1.498		\\
16227031	&			—			&		—		&		—		&		44,45,46		&		—		&		—		&		44		&		0.663	&		1.212		\\
17308656	&			—			&		—		&		—		&		—		&		4.9199$\pm$0.0035		&		0.58$\pm$0.08		&		43,44		&	—	&		2.794		\\
27858644	&		2.832 $\pm$ 0.002				&		0.17		&		26,40,54		&		53		&		2.84140 $\pm$ 0.00039		&		0.34$\pm$0.03		&		26,40		&		0.703	&		1.096		\\
32422219	&		2.112 $\pm$ 0.002				&		0.19		&		3,30		&		—		&		2.15012$\pm$0.00024		&		0.53$\pm$0.06		&		3,30		&		0.902	&		2.698		\\
34014829	&		3.859 $\pm$ 0.007				&		0.24		&		2,29		&		—		&		3.8632 $\pm$ 0.0016		&		0.44$\pm$0.03		&		2,29		&		0.892	&		5.417		\\
143029977	&		3.840 $\pm$ 0.014				&		0.1		&		31		&		4		&		3.8356 $\pm$0.0010		&		0.17$\pm$0.04		&		4,31		&		0.888	&		0.999		\\
193974787	&		3.072 $\pm$ 0.009				&		0.24		&		16		&		23,24,49,51		&		3.0942$\pm$0.0010		&		0.48$\pm$0.08		&		16,23,24		&		0.651	&		0.998		\\
198107795	&		11.292 $\pm$ 0.039				&		0.52		&		16,23,49,50		&		—		&		11.4334 $\pm$ 0.0058		&		0.59$\pm$0.03		&		16,23		&		0.845	&		11.011		\\
202408306	&		15.336 $\pm$ 0.153				&		0.76		&		23,50		&		—		&		15.34$\pm$0.03		&		0.66$\pm$0.06		&		23		&		0.993	&		0.946		\\
219095664	&		3.888 $\pm$ 0.001				&		0.24		&		14-21,24		&		22,23,40,41,		&		3.89442 $\pm$ 0.00041		&		0.18$\pm$0.04		&		14-25,40,41		&		0.852	&		4.316		\\
		&						&				&		25,55,57,58		&		47-52,54,56,59		&				&				&				&		&							\\
229653723	&			—			&		—		&		—		&		—		&		—		&		$<$0.34		&		22		&	—	&		1.231		\\
239097694	&			—			&		—		&		—		&		19,59		&		—		&		—		&		19		&		0.267	&		0.993		\\
286963145	&			—			&		—		&		—		&		23		&		—		&		$<$0.28		&		23		&		0.962	&		1.095		\\
288506050	&			—			&		—		&		—		&		16,22,23		&		2.4221 $\pm$ 0.0003		&		0.70$\pm$0.05		&		16,22,23		&		0.986	&		1.178		\\
307956653	&		3.614 $\pm$ 0.001				&		0.4		&		16-19,21,22,25,26,40		&		14,15,24,51		&		3.62703$\pm$0.00031		&		0.49$\pm$0.1		&		14-26,40,41		&		0.853	&		1.150		\\
		&		 				&				&	41,47-50,53,55-58,60		&		52,54,59		&				&				&					&		&							\\
311188315	&			—			&		—		&		—		&		24,51		&		1.95976$\pm$0.00043		&		0.43$\pm$0.08		&		24		&		0.372	&		1.661		\\
438883548	&			—			&		—		&		—		&		—		&		3.992$\pm$0.017		&		0.42$\pm$0.07		&		11		&	—	&		1.445		\\
441000085	&			—			&		—		&		—		&		11,38		&		3.9614 $\pm$ 0.0009		&		0.77$\pm$0.06		&		11,38		&		0.831	&		1.185		\\
902237947	&		6.000 $\pm$ 0.036				&		3.93		&		36		&		—		&		5.984$\pm$0.001		&		0.6$\pm$0.02		&		9		&		0.210	&		3.117		\\
\enddata
\end{deluxetable*}
}

\begin{figure*}[h!]
	\centering

    \includegraphics[scale=.6]{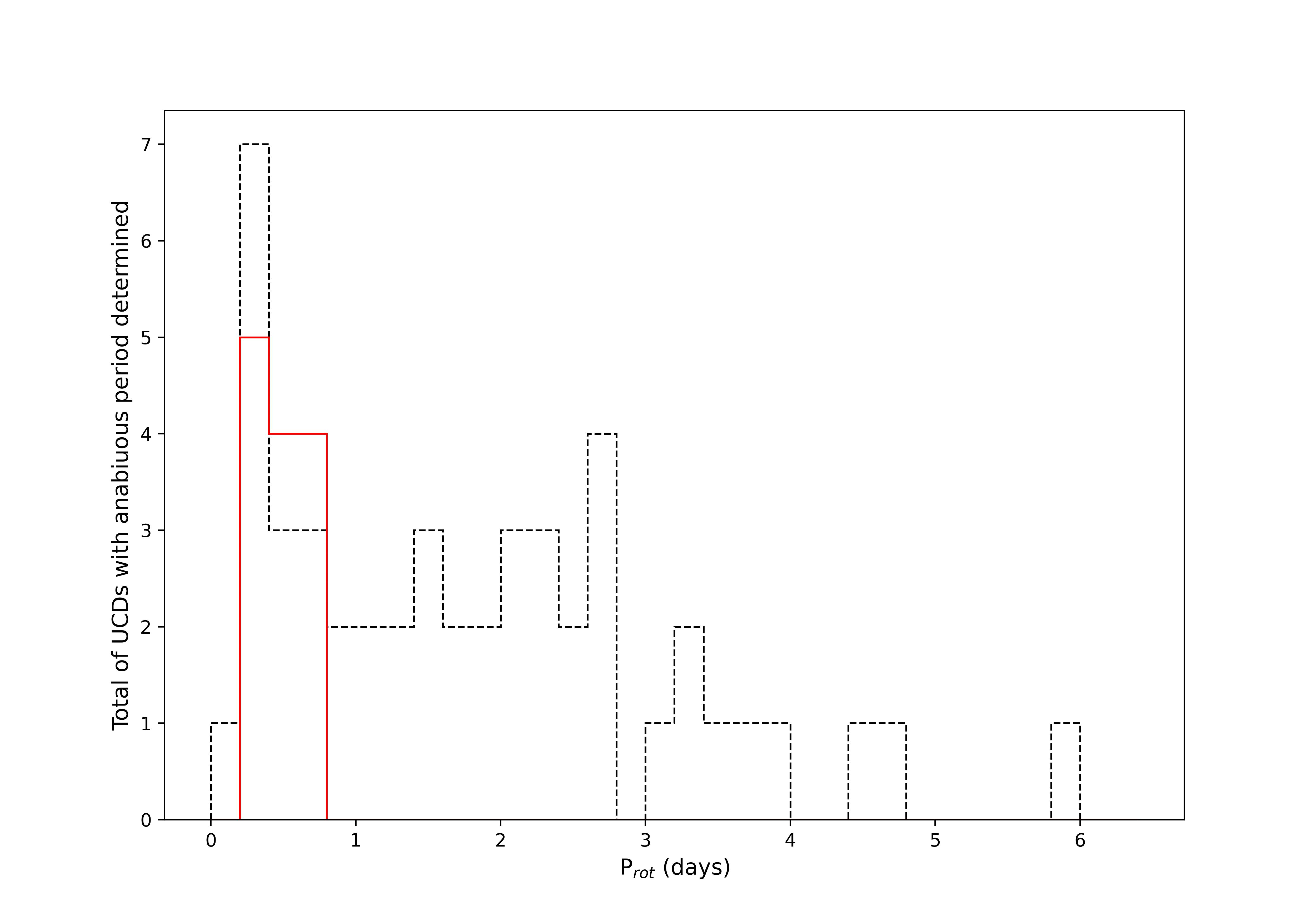}
	\caption{The distribution of rotation periods from the present analysis for the UCDs with rotation signature and single source behavior. The dashed black histogram stands for the  48 UCDs with unambiguous rotation periods, and the red color for the  13 UCDs withunambiguous rotation and low amplitude variability in the LCs.} 
	\label{Hist}
\end{figure*}

\begin{figure}[!ht]
	\centering
     \includegraphics[width=0.40\textwidth]{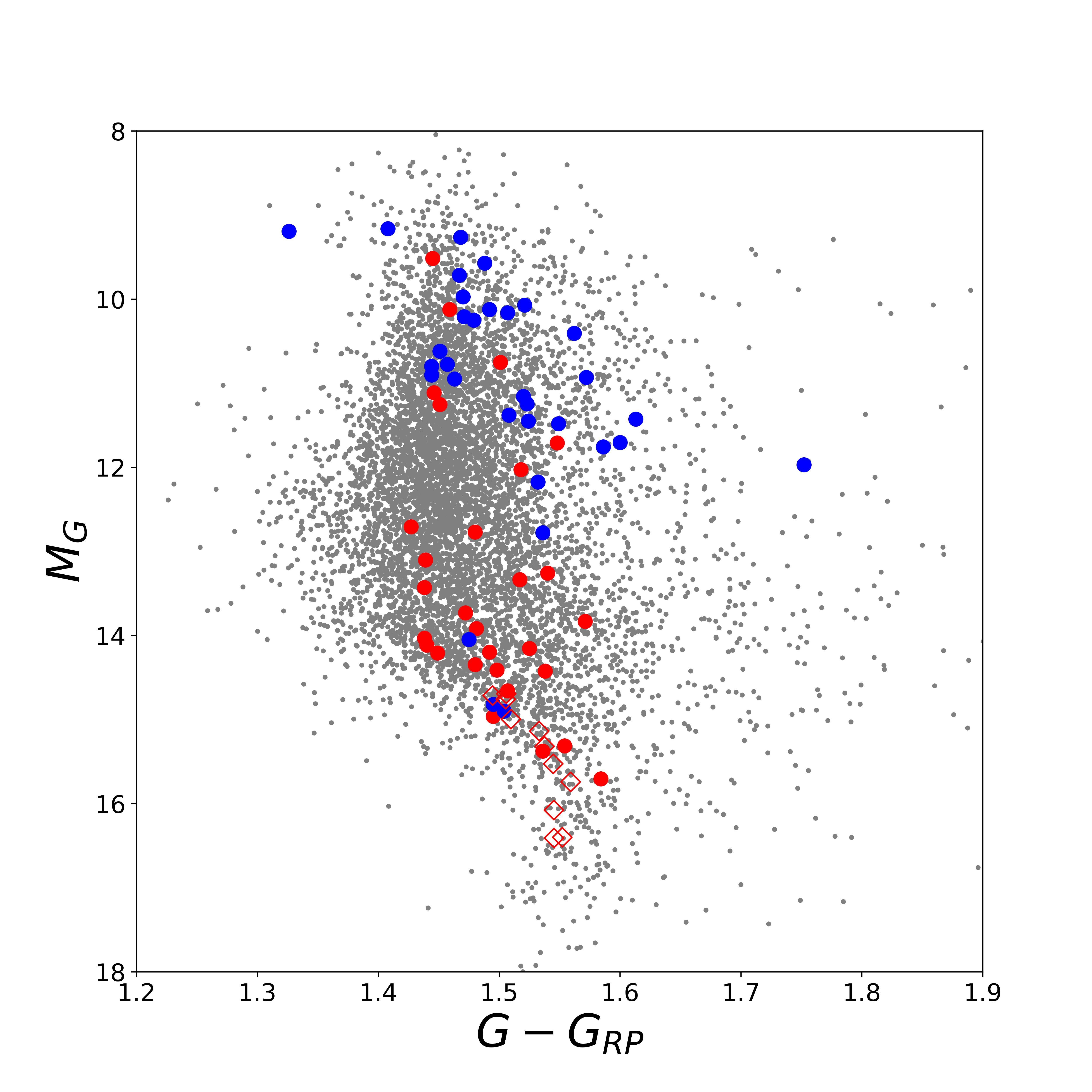}
     \includegraphics[width=0.50\textwidth,height=0.40\linewidth]{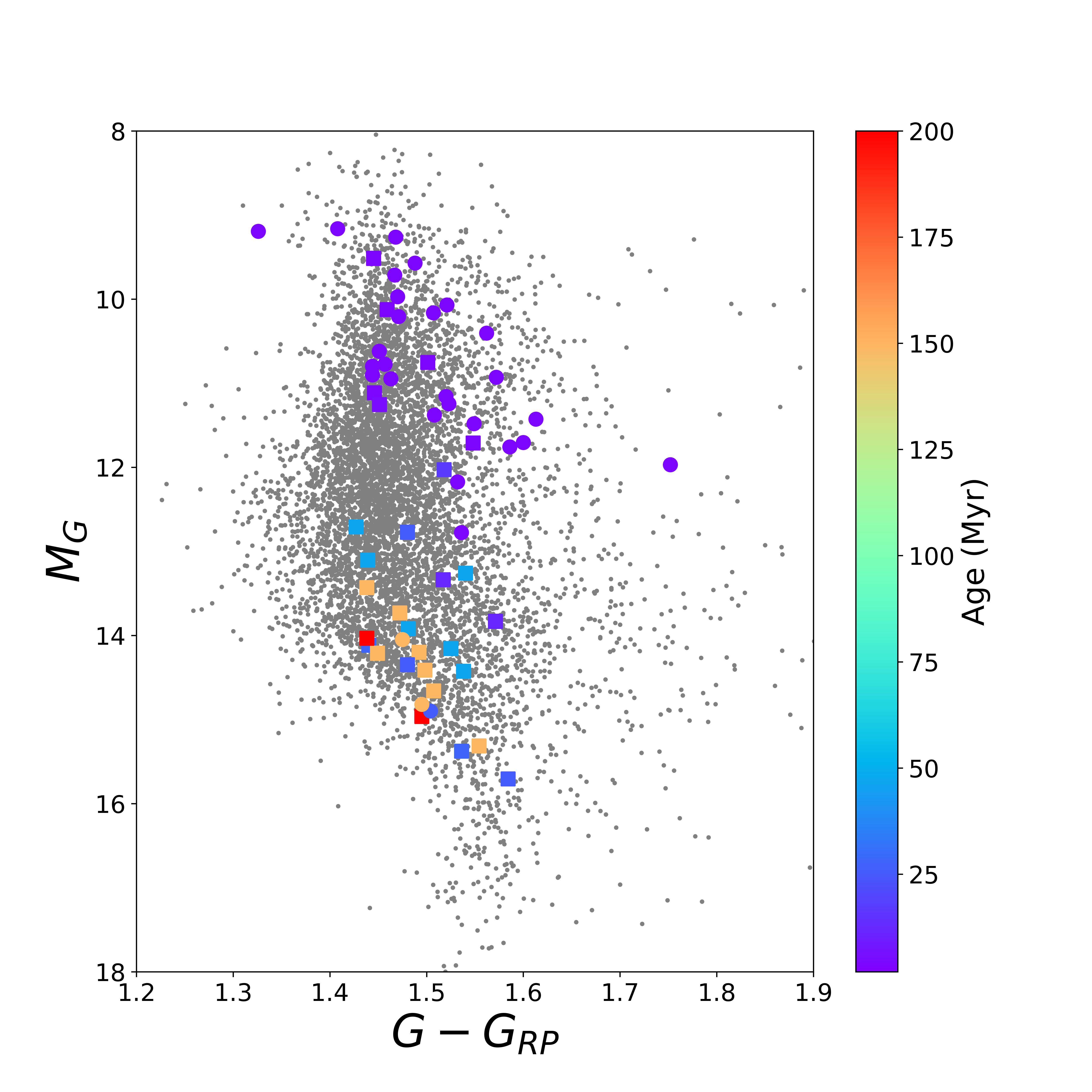}
	\caption{ The Gaia CMD for the sample of UCDs composing the present study, with black circles illustrating the whole sample of 7630 Gaia DR3 UCDs, given by \citet{Sarro2023}. In the left panel circles in blue are for the  UCDs with $P_{rot} \geq 1.0$ d,  red circles are for the  UCDs with P$_{rot}<1$ d, and red open diamonds stand for the UCDs listed by \citet{MilesPaez2023}. In the right panel, colors follow the color bar age intervals, which gives the age in Myr.}
	\label{CMD}

\end{figure}




\end{document}